\documentclass[runningheads]{llncs}
\usepackage[T1]{fontenc}
\usepackage{graphicx}
\usepackage{amsmath}
\usepackage{amssymb}
\usepackage{booktabs}
\usepackage{glossaries}
\usepackage{multirow}
\usepackage{placeins}
\usepackage{diagbox}
\usepackage{siunitx}
\usepackage[export]{adjustbox}
\usepackage{svg}
\usepackage{subfig}

\usepackage[capitalize]{cleveref}
\crefname{section}{Sec.}{Secs.}
\Crefname{section}{Section}{Sections}
\Crefname{table}{Table}{Tables}
\crefname{table}{Tab.}{Tabs.}

\usepackage{xspace}

\makeatletter
\DeclareRobustCommand\onedot{\futurelet\@let@token\@onedot}
\def\@onedot{\ifx\@let@token.\else.\null\fi\xspace}

\def\eg{\emph{e.g}\onedot} 
\def\ie{\emph{i.e}\onedot}

\def\etal{\emph{et al}\onedot}
\makeatother

\newacronym{enf}{ENF}{electric network frequency}
\newacronym{snr}{SNR}{signal-to-noise ratio}
\newacronym{rt}{RT}{reverberation time}
\newacronym{dl}{DL}{deep learning}
\newacronym{cnn}{CNN}{convolutional neural network}
\newacronym{crnn}{CRNN}{convolutional recurrent neural network}
\newacronym{rnn}{RNN}{recurrent neural network}
\newacronym{nn}{NN}{neural network}
\newacronym{seq2seq}{seq2seq}{sequence-to-sequence}
\newacronym{rir}{RIR}{room impulse response}
\newacronym{pp}{pp}{percentage points}

\begin{document}
\title{Towards Unconstrained Audio Splicing Detection and Localization with Neural Networks}
\author{Denise Moussa\inst{1,2,3}\orcidID{0000-0002-1390-9198} \and
Germans Hirsch\inst{2,3} \and
Christian Riess\inst{2}\orcidID{0000-0002-5556-5338}}
\authorrunning{D. Moussa et al.}
\titlerunning{Towards Unconstrained Audio Splicing Detection and Localization with NNs}
\institute{Federal Criminal Police Office (BKA), Germany \and
Friedrich-Alexander Universität Erlangen-Nürnberg, Germany \and Both authors contributed equally to this work.
\\
%\url{http://www.springer.com/gp/computer-science/lncs}
\email{\{denise.moussa, christian.riess\}@fau.de}}
\maketitle              % typeset the header of the contribution
\begin{abstract}
\setcounter{footnote}{0} 

%The abstract should briefly summarize the contents of the paper in
Freely available and easy-to-use audio editing tools make it straightforward to perform audio splicing.
Convincing forgeries can be created by combining various speech samples from the same person.
Detection of such splices is important both in the public sector when
considering misinformation, and in a legal context to verify the integrity of
evidence.
Unfortunately, most existing detection algorithms for audio splicing use
handcrafted features and make specific assumptions.
However, criminal investigators are often faced with audio samples from unconstrained sources with unknown characteristics, which raises the need for more generally applicable methods.

With this work, we aim to take a first step towards unconstrained audio splicing detection to address this need.
We simulate various attack scenarios in the form of post-processing operations that may disguise splicing.
We propose a Transformer \gls{seq2seq} network for splicing detection and localization. 
Our extensive evaluation shows that the proposed method outperforms existing
dedicated approaches for splicing detection~\cite{capoferri2020speech,jadhav2019audio} as well as the
general-purpose networks EfficientNet~\cite{tan2019efficientnet} and
RegNet~\cite{radosavovic2020designing}. Our source code is available at: \url{https://www.cs1.tf.fau.de/research/multimedia-security/code}
\end{abstract}

\section{Introduction}
With steadily improving technical methods, it became increasingly easier to convincingly manipulate multimedia content.
In the era of social media platforms, where huge amounts of images, audio snippets and videos are distributed, well-crafted pieces of maliciously manipulated content may quickly spread over the internet.
Forged multimedia content also becomes increasingly relevant in a legal context, such as criminal investigations.
Hereby, multimedia material may serve as clue or evidence.
Hence, it is of increasing importance to research and to deploy methods for ensuring the integrity and authenticity of such data.

Plausible multimedia forgeries can be created with several \gls{dl} methods. Examples are approaches for changing the attributes of a given image of a human face~\cite{viazovetskyi2020stylegan2}, for swapping the face of persons in photographs~\cite{nirkin2019fsgan}, or for face reenactment
~\cite{nirkin2019fsgan,thies2016face2face}. Another impressive demonstration was shown recently on video synthesis: given a speech recording and image of an individual's face, lip synthesis can even create a video of the individual that gives the speech~\cite{chen2019hierarchical}.
To create fake audio, voice conversion and speech synthesis have been proposed.
Voice conversion transforms the speech of one person to sound like the voice of another person~\cite{gao2018voice}.
Speech synthesis generates fully synthetic audio in the voice of a specific individual~\cite{jia2018transfer}.

In this work, we analyze audio splicing, which describes the processes of inserting, deleting or concatenating audio signals.
This tampering technique is arguably one of the easiest to perform, as
it can also be done by laypeople using freely available audio
editing tools like Audacity~\cite{audacity} or Oceanaudio~\cite{oceanaudio}.
Despite the simplicity of creation, audio splicing can pose a serious threat.
In some situations, even small changes can already alter the semantics of statements in a speech. For example, removing the negation particle `not' may flip a negated statement into a positive statement.  

Many existing works use hand-crafted features to extract specific characteristics for audio splicing detection or localization (\cref{subsec:manual_features}).
One weakness of hand-crafted features is that they are typically only applicable within relatively narrow bounds around their designed purpose.
For example, methods intended for finding differences in recording devices do not indicate splicing if the same device is employed. As a second example, classification of recording environments is unable to detect splicing from  static surroundings.
However, \gls{dl} techniques make it feasible to learn
features that fit more complex data distributions, and thereby to overcome the constraints of manual feature selection.
Surprisingly, \gls{dl} methods have so far barely been explored for audio
splicing detection.

With this work, we aim at closing this gap and propose a Transformer~\cite{vaswani2017attention} \gls{seq2seq} architecture to make a first step towards unconstrained audio splicing detection and localization.
We analyze different attack scenarios in the form of post-processing operations that may be applied to disguise splicing. This includes MP3 and AMR-NB single and multi compression, additive synthetic and real noise. Our forgery settings also cover audio compositions with samples from different environments, same environments and -- particularly difficult to differentiate -- anechoic environments.
We compare to a recent method that uses handcrafted
features~\cite{capoferri2020speech}, and to the neural network by Jadhav~\emph{et al.}~\cite{jadhav2019audio}, which to our knowledge is the only other end-to-end approach to forensic audio splicing detection.
Our method also outperforms the state-of-the-art general-purpose models EfficientNet~\cite{tan2019efficientnet} and RegNet~\cite{radosavovic2020designing}. Our main contributions are:
\begin{itemize}
	\item We present a robust end-to-end trainable \gls{seq2seq} model for audio splicing detection and localization.
	\item For evaluation, we propose a challenging data set that covers a large variety of forgery scenarios for the task of unconstrained audio splicing detection and localization.
\end{itemize}

\section{Related Work}
We distinguish two types of methods for audio splicing detection and localization. First, we review methods that manually extract features in \cref{subsec:manual_features}. Second, we review \gls{dl} methods in \cref{subsec:dl_methods}.

\subsection{Methods with Manual Feature Selection}
\label{subsec:manual_features}
Most related works on audio splicing detection rely on specific handcrafted
features.  As such, the feature response is inherently explainable, but the
methods can typically only be used in very specific
application scenarios. 
For example, Yang~\etal focus specifically on MP3 encoded audio~\cite{yang2008detecting}. They localize splicing in the signals by identifying inconsistencies in the offsets of the encoded audio frames.
Cooper~\etal target uncompressed audio, where they search discontinuities in the high frequencies of the audio waveform~\cite{cooper2010detecting}.

Some other approaches detect splicing by analyzing noise levels in the audio signal. For instance, Pan \etal compute global and local noise levels of audio data and identify abnormal changes in the noise level~\cite{pan2012detecting}. 
Meng \etal adopt a similar approach and compute local noise levels w.r.t. each syllable in the speech signal~\cite{meng2018detecting}.
Different from those works, Yan \etal~\cite{yan2021exposing} recently targeted composite audio with signals from different sources but with similar or equal \gls{snr}.
This work locates splicing points from the variances of Mel frequency cepstral coefficients (MFCCs).

Cuccovillo \etal~\cite{cuccovillo2013audio} perform splicing detection based on microphone classification. They exploit the fact that source files from different devices exhibit characteristic traces in the signal.

Several works analyze the \gls{enf}. Esquef \etal use \gls{enf} analysis to exclusively target splices in silent (non-voice-active) positions in the signal~\cite{esquef2015improved}.
However, \gls{enf} traces are weak and may be occluded by high noise.
Hence, Lin~\etal apply a spectral phase analysis instead which exhibits better robustness towards noise corruptions~\cite{lin2017exposing}.
Lin~\etal also present a second approach to increase the robustness to noise~\cite{lin2017supervised}. They propose to amplify abnormal changes 
via wavelet-filtering the \gls{enf} signal, and then extract autoregressive
coefficients for tampering detection. 

Another line of work models acoustic environmental signatures for splicing detection.
Zhao \etal compute features from the acoustic channel impulse response and ambient noise to model such an acoustic environmental signature~\cite{zhao2014audio,zhao2017audio}.
Rouniyar \etal refine the feature selection and increase performance by using both acoustic channel response and dynamic and static logarithmic spectral characteristics to identify and locate splicing~\cite{rouniyar2018channel}.

Recently, Capoferri \etal proposed the \gls{rt} as single forensic trace to detect
splicing~\cite{capoferri2020speech}.
The \gls{rt} is assumed to differ in different surroundings.
Hence, changes in the \gls{rt} across different temporal windows of the audio signal are then used to localize splicing.

A shared drawback of all these approaches is that they are restricted by their specific assumptions.
This includes specific audio compression formats or the absence of compression~\cite{yang2008detecting,cooper2010detecting}, differing recording devices in the forged audio~\cite{cuccovillo2013audio} and changing recording environments, like changing noise~\cite{pan2012detecting,meng2018detecting,yan2021exposing}, changing acoustic impulse~\cite{zhao2014audio,zhao2017audio,capoferri2020speech} or indicative \gls{enf} patterns~\cite{esquef2015improved,lin2017exposing,lin2017supervised}. 
% or more general, combined features to model changing environment signatures.

\subsection{Deep Learning Methods}
\label{subsec:dl_methods}
\gls{dl} methods were successfully applied to various tasks in audio forensics, including double compression detection~\cite{luo2014detecting}, audio recapture detection~\cite{luo2015audio}, speech presentation attack detection~\cite{korshunov2018use}, and fake speech detection~\cite{zhang2021fake}.
Surprisingly, the detection of audio splicing received little attention so far.
Mao~\etal~\cite{mao2020electric} target audio tampering (including splicing) with a \gls{cnn} classifier for detection but without localization.
The \gls{cnn} does not directly operate on the audio samples, but instead on pre-defined \gls{enf} features.
Zhang~\etal~\cite{zhang2022aslnet} use an encoder-decoder architecture based on VGG-16~\cite{simonyan2014very} to learn a segmentation mask for spliced audio samples.
This method can be seen as a very preliminary splicing detector whose design is dependent on quite restrictive assumptions. Most notably, the architecture only permits the detection of splicings from snippets with 1s duration, where at the same time each snippet is a random crop from a different speaker.

Jadhav \etal~\cite{jadhav2019audio} use the short-term Fourier transform of spliced audio signals directly as input to a shallow \gls{cnn} architecture.
They report that  their method is robust under added white Gaussian
noise and dynamic range compression.  However, this work also does not consider the
practically highly relevant case of splices from sources of the same speaker.
Additionally, the evaluation of the method is somewhat limited to a relatively
small, non-diverse test set.

In this work, we first explore the task of broadly applicable, robust audio splicing
detection and localization.
We propose a \gls{seq2seq} Transformer~\cite{vaswani2017attention} that
operates on various representations of audio signals.
Transformers excel in various natural language processing
tasks~\cite{otter2020survey} and efficiently exploit sequential context information which is also present in audio data. In addition, \gls{seq2seq} methods have the
specific benefit that, differently from CNN based methods~\cite{jadhav2019audio,zhang2022aslnet}, they can very naturally process sequences of arbitrary
input/output lengths.

Contrary to previous approaches~\cite{jadhav2019audio,zhang2022aslnet}, we validate our method on a challenging, diverse dataset.
We include single and multiple splices, as well as various post-processing operations that 
potentially disguise the splicing, such as multiple MP3 and AMR-NB compressions
and added synthetic and real noise.
Moreover, we consider forgeries that are created from audios that are recorded in different environments, as well as the more difficult case of identical environments. Additionally, we investigate the very challenging intersplicing scenario, where samples are assumed to be relatively anechoic and stem from the same recording in a static surrounding.
With this diversity we also aim at mitigating the application constraints of previous methods that use handcrafted features (\cref{subsec:manual_features}). 

\section{Methods}
This section consists of four parts.
In \cref{subsec:datagen} we describe our setup for data generation, in \cref{subsec:task} we formally define our task.
\cref{subsec:transf} summarizes the Transformer design~\cite{vaswani2017attention} and \cref{subsec:proposed_net} describes our proposed architecture.

\subsection{Data Generation Pipeline}
\label{subsec:datagen}
The data for our experiments is generated with base data from the ACE~\cite{eaton2016estimation} and the Hi-Fi TTS (TTS)~\cite{bakhturina21_interspeech} datasets.
For the ACE dataset, we use the same data split as Capoferri \etal~\cite{capoferri2020speech}, thus $65$ speech samples between \SI{1.28}{\second} and \SI{97}{\second} from $5$ female and  $9$ male speakers.
For the TTS dataset, we consider all $6$ female and $4$ male speakers and use the $10$ longest utterances per person reading different books.
The audio signals vary between \SI{7}{\second} and \SI{20}{\second}.
We separate speakers in training, validation and test pools.
For ACE, we take $10$ speakers for training, $2$ for validation, and reserve one female and one male speaker for test.
For experiments where we train on ACE, we take all speakers of TTS for testing.
When training on the latter, we split it into $7$ speakers for training, and for validation and test we use $2$ speakers (male and female) each. 

\begin{figure*}[t]
	\centering
	\includegraphics[width=\textwidth]{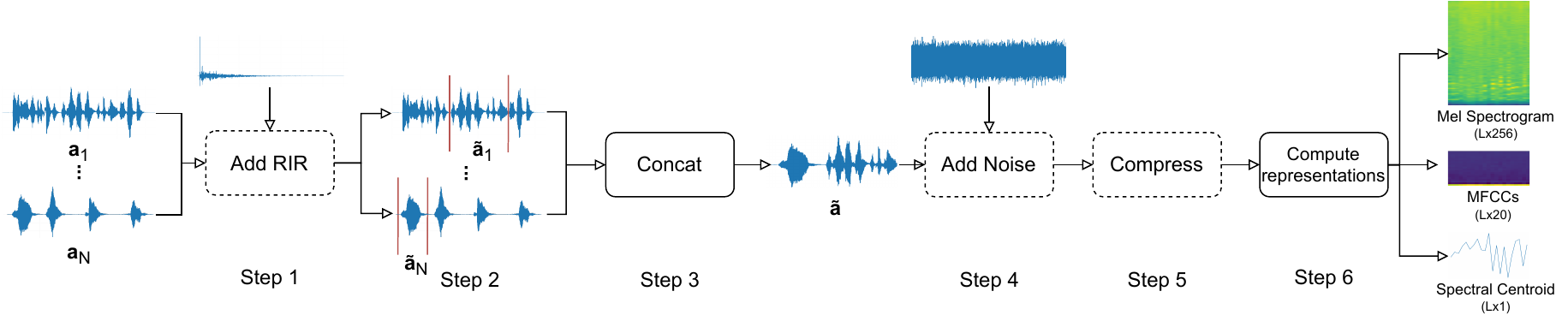}
	\caption{Adaptive pipeline for generating spliced audio samples. Reverberation of environments is simulated prior to splicing, while additive noise and compression are post-processing operations that may disguise splicing points. Dotted lines indicate optional operations, allowing for adaptable splicing forgery scenarios.}
	\label{fig:data_generation_pipeline}
\end{figure*}

For our experiments, we generate training, validation and test inputs from these disjoint speaker pools using our pipeline shown in~\cref{fig:data_generation_pipeline}. The pipeline operates as follows. First, we choose $N \in [1,5]$ audio signals $\mathbf{a}_1$(, \ldots, $\mathbf{a}_N)$ from the same speaker randomly and allow multiple choices of the same sample.
Each $\mathbf{a}_i$ is convolved with a \gls{rir} randomly sampled from a set of $7$ synthetic and $7$ real \glspl{rir} (step 1) to simulate recordings from environments with various echoic characteristics.
Here, we use the same \glspl{rir} as Capoferri \etal~\cite{capoferri2020speech} and allow recordings to stem from equal environments.
Note that we skip this step in one experiment without \gls{rir} cue (\cref{subsubsec:intersplicing}).
We omit any other individual processing per sample $\mathbf{a}_i$ (\eg, added
noise or compression) to avoid that our model is accidentally biased
to such overly specific features.
In the second step, we compute all silent, \ie non-voice-active, positions for each signal
with a voice activation detector. %\gls{vad}.
We sample  two random silent positions from each $\mathbf{a}_i$ and cut the sub-sequence between these points, which we denote as $\tilde{\mathbf{a}}_i$.
The spliced sample $\tilde{\mathbf{a}}$ is then the concatenation of all $\tilde{\mathbf{a}}_1, \ldots, \tilde{\mathbf{a}}_N$ or equal to $\tilde{\mathbf{a}}_1$ if $n =1$, \ie, no splicing was sampled (step 3).
Optionally, post-processing is applied to $\tilde{\mathbf{a}}$ which may mask splicing points.
To this end, additive noise can first be added with a specified \gls{snr} (step 4).
Then, compression can be applied once or multiple times with given format and strength (step 5).  The length of the resulting audio samples varies between \SI{3}{\second} and \SI{45}{\second}.
All specific post-processing parameters for each experiment are stated in \cref{sec:eval}.

As a last step, three representations of the resulting signal are computed via torchaudio~\cite{torchaudio} (step 6). We include the Mel spectrogram and MFCCs as two established features in the speech processing domain and additionally add spectral centroid features which describe the brightness of a sound~\cite{mckinney2003features}. Incorporating MFCCs and spectral centroid features additionally to the Mel spectrogram empirically showed to improve the performance in complex settings (\cref{sec:eval}).  For computing the spectral features, we use a window size of \SI{500}{\milli\second} and a sample rate of \SI{16}{\kilo\hertz}. For the Mel spectrogram we additionally choose $256$ Mel filter banks and then directly compute the MFCCs with $\texttt{n\_mfcc}=20$ from the result.
All other parameters are used in their default settings.

\subsection{Task Formulation}
\label{subsec:task}

We model audio splicing detection/localization as a \gls{seq2seq} task. The Mel Spectrogram of audio signals is our input sequence $\textbf{S} \in [-1,1]^{w\times H}$ with fixed $H$, \ie, fixed frequency resolution and variable width $w \in \mathbb{N}$ depending on the signal length in the time domain. We process $\textbf{S}$  as a series of slices, $\textbf{s}_i \in [-1,1]^H$, \ie, column-by-column, which yields the input series $\tilde{\textbf{S}} = [\textbf{s}_1, \textbf{s}_2, ..., \textbf{s}_w]$. 
Our specific task is to translate $\tilde{\textbf{S}}$ to a series of splicing points $\hat{P} = [\hat{p}_1, \hat{p}_2, ...,\hat{p}_{L}]$ of variable length $L$.  Each splicing location $\hat{p}_j \in \mathcal{V}$ is provided in seconds from a vocabulary $\mathcal{V}$ of all valid splicing locations. 

\subsection{Transformers}
\label{subsec:transf}
Transformers~\cite{vaswani2017attention} are \gls{nn} architectures with a scaled dot-product attention mechanism as core operation. They aim at grasping global dependencies between two sequences $\mathbf{S}_{i}$, $\mathbf{S}_{j}$. An important component is self-attention, where $\mathbf{S}_i = \mathbf{S}_j$.
The attention function maps a matrix triple $(\mathbf{Q}, \mathbf{V}, \mathbf{K})$  to an output, with 
$\mathbf{Q}$,$\mathbf{V}$,$\mathbf{K}$  
being a set of queries $\mathbf{Q}$, values $\mathbf{V}$, and keys $\mathbf{K}$. It is defined as
\begin{equation}
	\mathrm{Att}(\mathbf{Q},\mathbf{V},\mathbf{K}) = \mathrm{Softmax}\left(\frac{\mathbf{Q}\mathbf{K}^\top}{\sqrt{d_K}}\mathbf{V}\right),
\end{equation}
where $d_K$ is the dimension of the queries/keys that normalizes the product. $\mathbf{Q}, \mathbf{K}$ and $\mathbf{V}$ are obtained by projecting input sequences with learnable weight matrices $\mathbf{W}^Q \in \mathbb{R}^{d_m \times d_k}, \mathbf{W}^K \in \mathbb{R}^{d_m \times d_k}$ and $\mathbf{W}^V \in \mathbb{R}^{d_m \times d_v}$, where $d_m$ is the model dimension, \ie the output dimension of the model's layers, and $d_k, d_v$ are the keys' and values' dimensions, respectively.
%for input sequences. 
Typically, Transformers learn $h$ representations ($\mathbf{Q}, \mathbf{V}, \mathbf{K}$) and compute attention in parallel over those $h$ triples.
This is called multi-head attention and formally given by 
\begin{equation}
	\mathrm{MultiHead}(\mathbf{Q},\mathbf{V},\mathbf{K}) = [\mathrm{head}_{1} || ...|| \mathrm{head}_{h}]\mathbf{W}^O,
\end{equation}
where $\mathbf{W}^O \in \mathbb{R}^{hd_v \times d_m} $ and
$\mathrm{head}_i$ (for $1 \le i \le h$) is
\begin{equation}
	\mathrm{head}_{i} = \textrm{Att}(\mathbf{Q}_i,\mathbf{V}_i,\mathbf{K}_i)\enspace.
\end{equation}

A Transformer for \gls{seq2seq} tasks consists of a transformer encoder and a transformer decoder network.
The input sequence is processed by the encoder, which outputs the encoded result to the decoder.
The latter calculates the prediction from that encoder memory and the sequence elements from all previous time steps.
Both the encoder and the decoder network consist of several layers. Each encoder layer implements self-attention and fully-connected (FC) sub-layers.
Each sub-layer includes layer-normalization as well as a residual connection around itself.
The decoder is constructed similarly.
However, additionally to self-attention, it also computes the encoder-decoder attention, where $\mathbf{K}$, $\mathbf{V}$ stem from the encoder output, while $\mathbf{Q}$ stems from the decoder.
For an in-depth description of the Transformer architecture, we refer to Vaswani~\emph{et al.}~\cite{vaswani2017attention}.

\begin{figure}[t]
	\begin{center}
		\centering
		\includegraphics[width=0.98\textwidth]{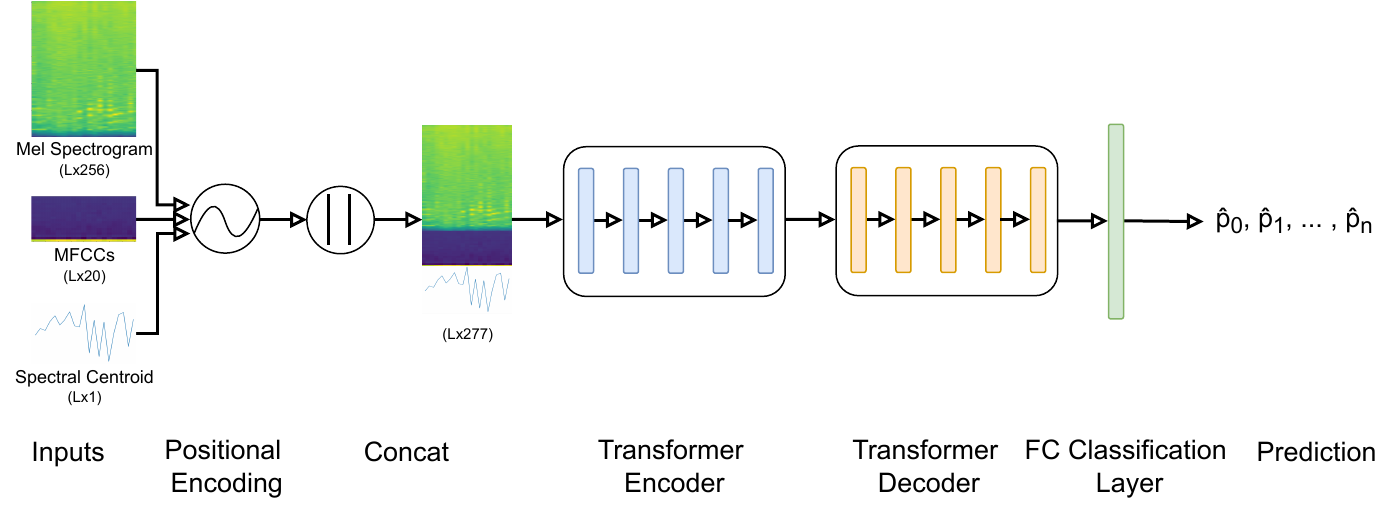}

	\end{center}
	\caption{Proposed network architecture. Different audio representations of length L (dependent on the audio sample length) are concatenated and processed by one encoder, which proved to be superior to separate encoders per representation. The encoded results are fed to the decoder and projected to the vocabulary space to yield the final predicted splicing points. }
	\label{fig:architecture}
\end{figure}
\label{subsec:architecture}

\subsection{Proposed Network Architecture}
\label{subsec:proposed_net}
Our proposed network architecture is depicted in \cref{fig:architecture}. It operates on three input representations, the Mel spectrogram, MFCCs and spectral centroid. All inputs are normalized to range $[-1,1]$ to support training stability. They are then subjected to positional encoding~\cite{vaswani2017attention}, concatenated and fed to the encoder (blue). Fusing the inputs and encoding them empirically showed to perform better than processing each input representation with a separate encoder and fusing the outputs. The encoder and decoder (orange) both consist of 5 layers with $8$ heads. The model dimension $d_m = d_k = d_v$ is of size $256$ and the dimension of each layer's FC sublayer is $512$. The last FC layer (green) projects the output of the decoder to the vocabulary size $|\mathcal{V}|$. Our vocabulary $\mathcal{V} =$ $\{\circ\}$ $\cup$ $\{0.5, ..., 44.5\}$ $\cup$ \{$<$pad$>$, $<$bos$>$, $<$eos$>$\} consists of $93$ items representing  no splicing ($\circ$), all possible $89$ splicing positions $[0.5, ...,44.5]$ in $500$\,ms steps as well as $3$ special tokens for \gls{seq2seq} translation.

\section{Evaluation}
\label{sec:eval}

We perform a variety of experiments to show the robustness and general applicability of our method. 

For single splicing, we report the top-$n$ accuracy (acc) per sample for $n \in [1,5]$. Thus, we report the relative frequency that all splicing points are predicted correctly within the $n$ most likely predictions.
To provide a more detailed insight on the quality of each prediction, we additionally report the Euclidean distance from the true splicing point $d_{\mathrm{sp}}$.
For multisplicing, we compute the Jaccard coefficient $\mathrm{J}$ between the predicted splicing points $\hat{P}_{\tilde{\mathbf{S}}}$  and ground truth  $P_{\tilde{\mathbf{S}}}$ per sample $\tilde{\mathbf{S}}$. 
It is calculated as
\begin{equation}
	\mathrm{J} = \frac{|P_{\tilde{\mathbf{S}}}| \cap |\hat{P}_{\tilde{\mathbf{S}}}|}{|P_{\tilde{\mathbf{S}}}| \cup |\hat{P}_{\tilde{\mathbf{S}}}|}\enspace,
\end{equation}
where $0$ means no intersection with the ground truth of (non)-splicing points and $1$ is the perfect score. Note that $\mathrm{J}$ is independent of the elements' order in ${P}_{\tilde{\mathbf{S}}}$ and $\hat{P}_{\tilde{\mathbf{S}}}$. We additionally report the recall $\mathrm{R}$. For both metrics, we tolerate time errors for splicing points within a narrow window of size $w=\SI{0.5}{\second}$. We also report softer variants $\mathrm{J}_w$ and $\mathrm{R}_w$ with larger tolerance windows of $w \in \{\SI{1}{\second}, \SI{2}{\second}, \SI{3}{\second}\}$.

\subsection{Models and Training Procedures}
\label{subsec:training}

We consider various baseline methods in our experiments. To the best of our
knowledge, 
the only end-to-end trainable \gls{dl} method for audio splicing detection was proposed by
Jadhav~\etal~\cite{jadhav2019audio}. We do not compare to the method by
Zhang~\etal~\cite{zhang2022aslnet} due to its very limiting constraints (cf. \cref{subsec:dl_methods}), which are not
straightforward to relax to our more general task.
We re-implemented the
method of Jadhav~\etal~\cite{jadhav2019audio} and filled unspecified network components with plausible settings.
This includes the padding strategy, the kernel size $k$ of the convolutional layers and the dimension $d$ of the two FC layers, which we chose as ``same'' padding, $k=3$ and $d=1024$. We additionally compare to state-of-the-art EfficientNet B0~\cite{tan2019efficientnet} (EffNet-B0) and RegNet-x-400mf~\cite{radosavovic2020designing} (RegNet-400m) recognition models~\cite{pytorchmodels}. All models, including our approach, are both evaluated with single-input, which includes only the Mel spectrogram as feature, as well as multi-input, which includes all three representations as shown in \cref{fig:data_generation_pipeline}.

Among all methods, our model is the most parameter-efficient with $\sim 7\text{M}$, followed by RegNet-400m with $\sim 29\text{M}$, EffNet-B0 with $\sim 71\text{M}$ parameters and the work by
Jadhav~\etal with $\sim 197\text{M}$ parameters.

We adapt the baseline \glspl{nn} to our task.
Contrary to our model that generalizes naturally to varying lengths, the \glspl{cnn} expect fixed input/output sizes.
We thus pad all inputs to the maximum length of $90$ and concatenate all three audio representations along the height. %where the Mel specrogram, MFCCs and spectral centroid are of the heights 256, 20 and 1, respectively. 
The input layer is thus set to $1\times90\times277$.
For multisplicing, we extend the output FC layer from one to the maximum possible number $N$ of splicing positions, predicting $\hat{p}_i \in [0.5, 1.0, \ldots, 44.5]$  with $i \in [1,N]$ or special token $\hat{p}_{N+1} = \circ$ if no splicing is detected. Given the corresponding ground truth, all models, including ours, are trained using the cross-entropy loss.

Per model, we increase the computational speed by enlarging the training batch
size for as long as the training still convergences.
We chose a batch size of $512$ for our model, $256$ for Jadhav \etal~\cite{jadhav2019audio} and RegNet-400m~\cite{radosavovic2020designing}, and $64$ for EffNet-B0~\cite{tan2019efficientnet}.
Still, the Adam optimizer with learning rate
$1\mathrm{e}^{-4}$ proved to be most beneficial for all models.
We train with early stopping on the validation loss with delta $\delta = 0.2$
and a patience of $10$ epochs. Our model is trained with teacher forcing.

We also compare our method to Capoferri \etal~\cite{capoferri2020speech}, as a
recently proposed method with handcrafted features for single splicing detection. For the first part of the comparative evaluation in \cref{subsec:single_splicing}, we include \glspl{rir} in step 1 of our data generation pipeline (cf.\ \cref{fig:data_generation_pipeline}). Hence, this feature is
present in our data and their analytic approach is applicable in this case. Note, that we consistently allow the same \gls{rir} to be sampled for audio sources to model both equal and different environmental sources.

\subsection{Basic Single-Splicing Forgery Model}
\label{subsec:single_splicing}
\newcommand{\iheight}{2.9cm}
\newcommand{\crossheight}{2.5cm}
In our first experiments, we consider basic forgery scenarios that only cover single splices. We first conduct a  cross-dataset validation (\cref{subsec:cross_data}), where we omit any post-processing operations. In a second step, the scenario is more difficult with %evaluate the robustness and adaptability towards 
additive synthetic noise and single compression post-processing (\cref{subsec:noise_comp}) as easy mechanisms to disguise splicing.

\subsubsection{Cross-Dataset Validation}
\label{subsec:cross_data}

We perform a cross-dataset validation experiment between data from the ACE~\cite{eaton2016estimation} and the TTS~\cite{bakhturina21_interspeech} dataset.
For each, we generate a training set of $500\:000$ and multiple test sets of $30\:000$ samples as described in \cref{subsec:datagen}.
For now, we omit post-processing operations (step 4,5) in the generation pipeline (cf.\ \cref{fig:data_generation_pipeline}). This corresponds to the most basic forgery model, where new content is generated from samples of the same speaker with no camouflage operations to disguise the splicing points.

\begin{figure}[t]
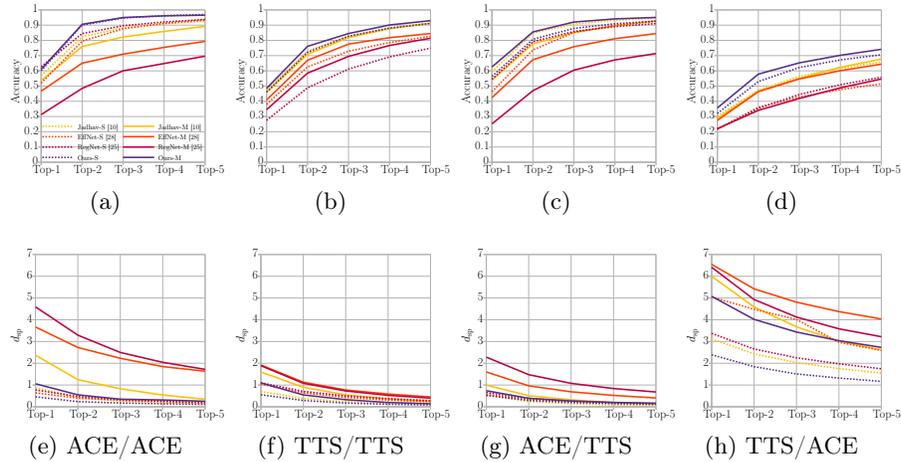

	
	\subfloat[\label{subfig:ace-acc}]{
		\scalebox{0.26}{
			\includesvg{images/cross/ace-acc.svg}}}
		\subfloat[\label{subfig:tts-acc}]{
		\scalebox{0.26}{
			\includesvg{images/cross/tts-acc.svg}}}
		\subfloat[\label{subfig:acetts-acc}]{
		\scalebox{0.26}{
			\includesvg{images/cross/acetts-acc.svg}}}
		\subfloat[\label{subfig:ttsace-acc}]{
			\scalebox{0.26}{
				\includesvg{images/cross/ttsace-acc.svg}}}\\
	\subfloat[ACE/ACE\label{subfig:ace-dist}]{
		\scalebox{0.26}{
			\includesvg{images/cross/ace-dist.svg}}}
	\subfloat[TTS/TTS\label{subfig:tts-dist}]{
		\scalebox{0.26}{
			\includesvg{images/cross/tts-dist.svg}}}
	\subfloat[ACE/TTS\label{subfig:acetts-dist}]{
		\scalebox{0.26}{
			\includesvg{images/cross/acetts-dist.svg}}}
	\subfloat[TTS/ACE\label{subfig:ttsace-dist}]{
		\scalebox{0.26}{
			\includesvg{images/cross/ttsace-dist.svg}}}
	\caption{Results for cross-data validation for the single- (dotted lines) and multi-input (solid lines) variants of the models. Each column portrays one train set/test set run. Top row: top-$n$ accuracies. Bottom row: $d_{\mathrm{sp}}$. We perform best for all combinations. Generalizing from TTS shows to be the most difficult.}
	\label{fig:cross_dataset}
\end{figure}

\Cref{fig:cross_dataset} shows the results. The top row reports accuracy, the bottom row the distance $d_{\mathrm{sp}}$ to the ground truth splicing point. The four plots in one row show the four combinations of source datasets for training and testing.

Our method (violet) performs best in every case, while Jadhav~\etal (yellow) is the best baseline. In general, our method in multi-input mode (solid lines) achieves an equal (\cref{subfig:ace-acc}) or higher accuracy (\cref{subfig:tts-acc}-\ref{subfig:ttsace-acc}) than the single-input variant (dotted lines), even if $d_{\mathrm{sp}}$ is slightly larger (\cref{subfig:ace-dist}-\ref{subfig:ttsace-dist}).
Contrary, the CNN baselines~\cite{jadhav2019audio,radosavovic2020designing,tan2019efficientnet} mostly perform worse when in multi-input mode.
We assume a slight over adaptation due to the additional input representation information, such that training on the single (sparser) representation is beneficial in this case. 
Concerning the cross-dataset combinations, generalizing from TTS to ACE is the hardest for all models (\cref{subfig:ttsace-acc}, \cref{subfig:ttsace-dist}), while both the ACE/TTS (\cref{subfig:acetts-acc}, \cref{subfig:acetts-dist}) and the intra dataset experiment on TTS (\cref{subfig:tts-acc}, \cref{subfig:tts-dist}) exhibit significantly better results. We thus assume that TTS may contain specific characteristics to which the models adapt to, which results in a decreasing performance when testing on another dataset.
For this reason, we take ACE/TTS as training/test combination for all following experiments.

We also run the method by Capoferri~\etal~\cite{capoferri2020speech} on both datasets.
By design, it only provides top-1 accuracies.
We report an accuracy of $4.9\%$ and $6.8\%$, as well as a $d_{\mathrm{sp}}$ of $4.60$ and $2.93$ for ACE and TTS, which is below the other methods.
We hypothesize that the discrepancy to the reported performances by Capoferri~\etal is due to differences in the dataset construction.
While Capoferri \etal~\cite{capoferri2020speech} randomly splice signals, we only splice during silent points, which makes the detection task considerably more difficult.

\subsubsection{Influence of Noise and Compression}
\label{subsec:noise_comp}

\begin{figure}[t]
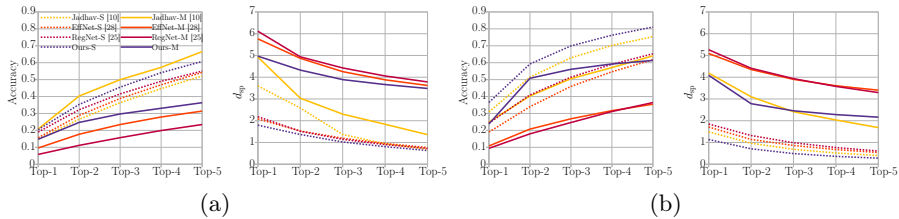

	\subfloat[\label{subfig:rob}]{
		\scalebox{0.26}{
			\includesvg{images/rob-adapt/acc-rob.svg}}
			\scalebox{0.26}{
			\includesvg{images/rob-adapt/dist-rob.svg}}}
	\subfloat[\label{subfig:adapt}]{
		\scalebox{0.26}{
			\includesvg{images/rob-adapt/acc-adapt.svg}}
			\scalebox{0.26}{
			\includesvg{images/rob-adapt/dist-adapt.svg}}}
	\caption{Results for robustness (\cref{subfig:rob}) and adaptability tests (\cref{subfig:adapt}) for single- (dotted lines) and multi-input (solid lines) models. 
 Overall, our model adapts best (\cref{subfig:adapt}) and is second best w.r.t
 robustness (\cref{subfig:rob}). Contrary to most other experiments, the
 single-input accuracy is higher than multi-input
 accuracy.}
	\label{fig:rob-adapt}
\end{figure}

To disguise splicing points, a forger may take simple measures. Here, we consider additive synthetic noise, MP3 compression, and AMR-NB compression as post-processing operations on the forgery. Additive noise may be easily added with openly available audio editing tools~\cite{audacity,oceanaudio}. Encoding the audio data in a lossy compression format is even more straightforward.
We consider white Gaussian noise as a general noise approximation and MP3 and AMR-NB as two popular lossy compression formats.
We generate a test set from the TTS test split (\cref{subsec:datagen}) with $30\:000$ samples and consider weak to strong degradation influence (step 4,5 in \cref{fig:data_generation_pipeline}). In detail, per sample, we first randomly chose a \gls{snr} $\in [-10, 50]\,\textrm{db}$ for noise. Then, randomly either MP3 or AMR-NB is chosen for compression with random $b_r \in [10, 128]\,\frac{kb}{s}$ for MP3 and $b_r \in \{4.75,$ $5.15,$ $5.9,$ $6.7,$ $7.4,$ $7.95,$ $10.2,$ $12.2\}\,\frac{kb}{s}$ for AMR-NB. In addition, we evaluate the adaptability of the models towards all degradations. For this, we fine-tune all \glspl{nn} on a ACE training set of $500\:000$ samples with the same degradation pipeline and rerun on the TTS test set.

The \glspl{nn}' results are reported in \cref{fig:rob-adapt}. \Cref{subfig:rob} shows the robustness of the models trained on the clean training set towards the degraded test set.
With the exception of Jadhav~\etal~\cite{jadhav2019audio}, the single-input models exhibit higher accuracy (\cref{subfig:rob}, left).
Our single-input model performs second best w.r.t accuracy and has the lowest $d_{\mathrm{sp}}$ (\cref{subfig:rob}, right).
When fine-tuned on the degradations, all models perform better as expected (\cref{subfig:adapt}).
Similar to the CNNs in the cross-dataset setting, the single-input variants are superior, with the difference that also our approach performs better here with single-input.
This is, however, a transient effect: the multi-input variant consistently
performs better for more complex training and test settings as will be shown in
\cref{subsec:multi_splicing}. 

We also evaluate the handcrafted baseline by
Capoferri~\etal~\cite{capoferri2020speech}. Its performance is below the
\glspl{nn} for all experiments with an accuracy of $7\%$ and $d_{\mathrm{sp}} = 2.9$.
More in detail, the method achieves a good $d_{\mathrm{sp}} \leq 1.5$ for $30\%$ of the test samples. However, for the remaining $70\%$, the error is considerably larger.

\subsection{Advanced Multi-Splicing Forgery Model}
\label{subsec:multi_splicing} 
In our following experiments, we extend our forgery model to incorporate various multi-splicing scenarios for evaluating the robustness of the proposed method. To this end, we generate a training set of $500\:000$ samples from the ACE training split and fine tune the models previously trained on single splicing on it. We include $n \in [0, 5]$ splicing points in equal numbers, and apply all available processing steps, \ie, \glspl{rir}, additive white Gaussian noise, and single MP3 or  AMR-NB compression with the same \gls{snr} and $b_r$ ranges as in \cref{subsec:noise_comp}. 

For testing, we now assume a more elaborate forger and hence define several multi-splicing scenarios as they might be expected in the real world.
For each scenario, we use test sets with $10\:000$ samples from the TTS test speaker pool and test the models' generalization ability to the diverse splicing situations that differ from the training setting.
The method by Capoferri~\etal does not support multi splicing, hence we exclude it from this evaluation.

\subsubsection{Multi Compression}
\label{subsubsec:multicompression}
In this experiment, we consider the difficult case of a forger who not only masks multiple splicing points with noise, but who also compresses the result multiple times afterwards. Note that multi compression may also occur during recurring up- and downloading from the internet.

We consider $n_c \in [0,5]$ compression runs, which also includes uncompressed and single compressed files for reference. We generate $6$ test sets, one for each $n_c$.
The full data generation pipeline is applied with the same distortion parameters as for training, which includes RIRs as well as synthetic noise addition. However, the compression step is repeated $n_c$ times with randomly sampled parameters.

\begin{figure}
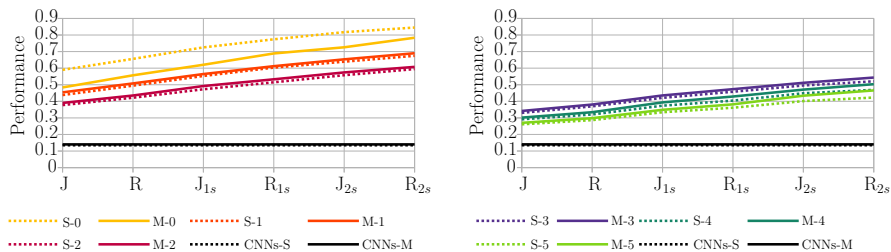


	\scalebox{0.425}{
	\includesvg{images/multicompression/multi1.svg}}
	\scalebox{0.425}{
	\includesvg{images/multicompression/multi2.svg}}
	\caption{Performance on $n_c \in [0,2]$ (left) and $n_c \in [3,5]$ (right) compression runs for our method trained on single- (dotted lines) and multiple-input (solid lines) representations. All baseline CNNs~\cite{jadhav2019audio,radosavovic2020designing,tan2019efficientnet} have difficulties to generalize to the task and perform equally with minimal fluctuations. Our multi-input model exhibits greater robustness than the single-input variant except for $n_c=0$.}
	\label{fig:multicompression}
\end{figure}

The evaluation results are reported in \cref{fig:multicompression}. All baseline \glspl{cnn}~\cite{jadhav2019audio,radosavovic2020designing,tan2019efficientnet} perform comparably and fluctuate around $13.5\%$ for single-input and $14\%$ for multi-input along all $n_c$. We note that they show generalization problems towards this task and mostly collapse to predicting the same splicing points for most samples. We ran a series of experiments to tune the hyperparameters of these methods, but were not able to resolve this problem. Since our model did not show such problems, we attribute the advantage to the Transformer \gls{seq2seq} design which is more naturally suited to the processing of sequential audio information than the per-position baseline classifiers.

As expected, the performance of our model decreases with increasing number of compression runs. Our multi-input method outperforms the single-input method except for the case of no compression, where the latter has a notable advantage of up to $10.0$ \gls{pp} for all metrics. For $w=\SI{1}{\second},\SI{2}{\second}$, the multi-input model improves further by $5.7$\gls{pp} for $\mathrm{J}_{1s}$ and $\mathrm{R}_{1s}$, and even by $10.7$\gls{pp} for $\mathrm{J}_{2s}$ and $10.5$\gls{pp} for $\mathrm{R}_{2s}$.
 
For the most difficult case with $n_c =5$ compression runs, our model still achieves $\mathrm{J_{2s}} =  43.4\%$ and $\mathrm{R_{2s}} = 46.6\%$ which we consider a notable generalization ability towards this difficult setting.

\subsubsection{Intersplicing}
\label{subsubsec:intersplicing}
Intersplicing addresses a particularly difficult scenario, where a forgery is spliced together from 
a speaker recorded in a static (\ie, unchanging) environment without significant impulse response cues. This can be expected from rather anechoic surroundings like open spaces. We omit \glspl{rir} and any further post-processing operations in this experiment and generate one test set for each number of $n
\in [0,5]$ possible splicing points. Due to little cues and strong deviation from the training set distribution, this experiment requires particularly strong generalization abilities.

In fact, similarly to the phenomenon described in \cref{subsubsec:real_world_noise}, no baseline CNN~\cite{jadhav2019audio,radosavovic2020designing,tan2019efficientnet} can meet this requirement. All baseline methods default to a constant performance of $16.7\%$ for $J_w$ and $R_w$ with $w \in \{\SI{0.5}{\second}, \SI{1}{\second}, \SI{2}{\second}\}$.
Both our single and multi-input model are able to generalize to a certain degree.
However, intersplicing proves to be the most challenging of all tested scenarios.
The single-input variant achieves a $\mathrm{J}_w$ of $17.2\%, 19.0\%$ and $21.3\%$  and a $\mathrm{R}_w$ of $17.9\%, 20.0\%$ and $22.7\%$ for $w \in \{\SI{0.5}{\second}, \SI{1}{\second}, \SI{2}{\second}\}$.
The multi-input variant performs comparably with a  $\mathrm{J}_w$ of $16.7\%, 17.5\%$ and $18.8\%$, and a   $\mathrm{R}_w$ of $17.0\%, 18.0\%$ and $19.4\%$ for $w \in \{\SI{0.5}{\second}, \SI{1}{\second}, \SI{2}{\second}\}$, respectively.
 
 Overall, we report that our Transformer \gls{seq2seq} is the only model that generalizes towards this task. However, there is still room for robustness improvements towards data with such large deviations from the training distribution.

\subsubsection{Real World Noise}
\label{subsubsec:real_world_noise}
In training, we cover additive white Gaussian noise to approximate background noises.
We now evaluate the robustness towards composite samples distorted by a forger with additive real noise to hide splicing points more convincingly. Real noise samples can be easily downloaded from the internet.
For our experiments, we chose free ambiance sound samples featuring rain\footnote{{\url{https://freesound.org/people/straget/sounds/531947/}}}, a train passing by\footnote{\url{https://freesound.org/people/theplax/sounds/615849/}}, an crowded exhibition hall\footnote{\url{https://freesound.org/people/BockelSound/sounds/487600/}}, and a crowded boarding gate at the airport\footnote{\url{https://freesound.org/people/arnaud\%20coutancier/sounds/424362/}}. We generate one test set per noise type with \glspl{snr} uniformly drawn from the range $[-10, 50]\,\textrm{db}$ per sample.  We also include \glspl{rir} and single compression post-processing as described in \cref{subsec:multi_splicing}.

\begin{figure}[t]
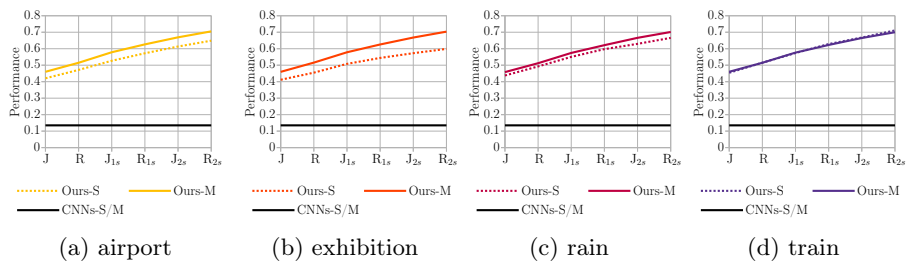

	\subfloat[airport\label{subfig:airport}]{
		\scalebox{0.35}{
			\includesvg{images/real_noise/airport.svg}
		}
	}
	\subfloat[exhibition\label{subfig:exhibition}]{
	\scalebox{0.35}{
	\includesvg{images/real_noise/exhibition.svg}}}
	\subfloat[rain]{
	\scalebox{0.35}{
	\includesvg{images/real_noise/rain.svg}}}
	\subfloat[train]{
	\scalebox{0.35}{
	\includesvg{images/real_noise/train.svg}}}
	\caption{Performance on $4$ real world noise types. The CNN baselines~\cite{jadhav2019audio,radosavovic2020designing,tan2019efficientnet} are unable to generalize. Our multi-input model (solid lines) is best except for train noise, where the single-input variant (dotted lines) performs comparably.}
	\label{fig:real_world_noise}
\end{figure}

The results are reported in \cref{fig:real_world_noise}. As in all previous multi-splice experiments, the tested CNNs~\cite{jadhav2019audio,radosavovic2020designing,tan2019efficientnet} are unable to generalize and yield a performance of $13.5\%$ consistently for all metrics. Our multi-input model outperforms the single-input variant especially for the more complex airport and exhibition background noise (\cref{subfig:airport}, \cref{subfig:exhibition}). For the former noise, the average increase over all  $w \in \{\SI{0.5}{\second}, \SI{1}{\second}, \SI{2}{\second}\}$  is $4.9$ and $5.3$\gls{pp} for $\mathrm{J}_w$ and $\mathrm{R}_w$, respectively. For the latter noise, the average increase is even $7.1$ and $8.2$\gls{pp}, respectively. Averaged over all real world noises, our best performing multi-input model shows satisfying robustness with $\mathrm{J} = 45.9\%$ and $\mathrm{R} = 51.4\%$ and for the $\SI{2}{\second}$ windows even reaches a performance of $\mathrm{J_{2s}}=66.7\%$ and $\mathrm{R_{2s}} = 70.3\%$.

\section{Conclusion}
This work investigates the robust detection and localization of single and multiple splices in audio forgeries under unconstrained settings.
We propose a Transformer \gls{seq2seq} network for this task.
We perform extensive evaluations that cover basic and advanced forgery models, including splicing of samples from different/same, echoic/anechoic recording surroundings, and post-processing operations like additive synthetic/real noise and single/multiple compression runs that may disguise splicing.
Our method clearly outperforms competing networks and CNN baselines, while requiring the smallest
number of parameters.
By design, it is also more universally applicable than methods with handcrafted features that exploit specific manipulation characteristics.

The proposed method generalizes well to challenging scenarios with multiple splices that cannot be solved by other CNNs.
We hypothesize that this is due to the better suitability of a \gls{seq2seq} model for processing sequential audio data.

After this first step into the direction of unconstrained audio splicing detection and localization we aim at further improving the robustness of our method.
We expect that there is still room for improvement in particularly challenging situations like the generalization to intersplicing (\cref{subsubsec:intersplicing})  or a large number of post-processing compression steps (\cref{subsubsec:multicompression}.)

 \bibliographystyle{splncs04}
\bibliography{refs}
\end{document}